\documentstyle[11pt]{article}

\input{epsf}

\newbox\grsign \setbox\grsign=\hbox{$>$} \newdimen\grdimen\grdimen=\ht\grsign \newbox\simlessbox \newbox\simgreatbox
\setbox\simgreatbox=\hbox{\raise.5ex\hbox{$>$}\llap
     {\lower.5ex\hbox{$\sim$}}}\ht1=\grdimen\dp1=0pt
\setbox\simlessbox=\hbox{\raise.5ex\hbox{$<$}\llap
     {\lower.5ex\hbox{$\sim$}}}\ht2=\grdimen\dp2=0pt
\def\simgt{\mathrel{\copy\simgreatbox}}

\begin{document}
\noindent
\hspace{5in} CAMK preprint 324

\noindent
\hspace{5in} August 1997

\vspace{1.7in}

\begin{center}
{\bf EVOLUTION OF PEAKS\\ 
IN WEAKLY NONLINEAR DENSITY FIELD\\ 
AND DARK HALO PROFILES}
\end{center}
\vspace{0.07in}
\begin{center}
{\sc Ewa L. \L okas}
\end{center}
\vspace{0.07in}
\begin{center}
{\em Copernicus Astronomical Center\\ 
Bartycka 18, 00-716 Warsaw, Poland \\
E--mail: lokas@camk.edu.pl}
\end{center}

\vspace{0.4in}

\begin{center}
{\bf ABSTRACT}
\end{center}

\noindent
Using the two-point Edgeworth series up to second order we construct the 
weakly nonlinear conditional probability distribution function for the 
density field around an overdense region. This requires calculating the 
two-point analogues of the skewness parameter $S_{3}$. We test the 
dependence of the two-point skewness on distance from the peak for 
scale-free power spectra and Gaussian smoothing. The statistical features 
of such conditional distribution are given as the values obtained within 
linear theory corrected by the terms that arise due to weakly nonlinear 
evolution. The expected density around the peak is found to be always 
below the linear prediction while its rms fluctuation is always larger 
than in the linear case. We apply these results to the spherical model of 
collapse as developed by Hoffman \& Shaham (1985) and find that in general 
the effect of weakly nonlinear interactions is to decrease the scale from 
which a peak gathers mass and therefore also the mass itself. In the case 
of open universe this results in steepening of the final profile of the 
virialized protoobject.


\begin{center}
Submitted to MNRAS
\end{center}

\newpage

\section{Introduction}

The simplest deterministic model of structure formation proposed by 
Gunn \& Gott (1972), called the spherical model, described the evolution 
of an overdense region in the otherwise unperturbed, expanding Universe.
It was extended by Gott (1975) and Gunn (1977) to apply to the evolution 
of matter around an already collapsed density perturbation superposed on a 
homogeneous background. The main prediction of the model (called the 
spherical accretion or the secondary infall model) was that the matter 
collapsing around the perturbation should form a halo with $r^{-9/4}$ 
density profile.

It is much more realistic to assume that the progenitors of structure were
not the collapsed perturbations but the local maxima (rare events) in the
density field which had initially Gaussian probability distribution. This 
was the approach of Hoffman \& Shaham (1985) (hereafter HS) who applied the 
secondary infall mechanism to the hierarchical clustering model. They 
assumed that the density peak dominates to some extent the surroundings 
causing the collapse of the material that is gravitationally bound to it. 
The initial density profile around the peak was approximated by the mean 
density predicted by the initial probability distribution which was 
assumed to be Gaussian. Thus a link was established between the 
statistical nature of the fluctuations and the deterministic character of 
the spherical model. HS considered scale-free initial power spectra of 
fluctuations and found that the final profiles of halos depend on the 
spectral index $n$ and on the density parameter $\Omega$.

One of the key assumptions underlying the calculations of HS was that the 
matter influenced by the peak collapses onto it undisturbed by the 
background. This is equivalent to the statement that the peak identified 
with some resolution (smoothing) scale collapses while the surrounding 
density field is still linear i.e. its rms fluctuation at this scale is 
much smaller than unity. Since the rms fluctuation grows with time and 
decreases with the smoothing scale, this approximation might be true for 
very early stages of evolution or very large smoothing scales. The example 
of the Virgo supercluster (which has not yet collapsed) shows that this is 
not always the case: assuming the power spectrum $P(k) \propto k^{-3/2}$ 
and knowing that at present the rms fluctuation is of the order of unity 
at the smoothing scale of 10 Mpc, the size of the supercluster being 
around 30 Mpc, one can easily estimate that the rms fluctuation at the 
scale of the supercluster is well in the weakly nonlinear regime.

The purpose of this paper is to present a generalization of the 
calculations of HS to the case of density peaks collapsing in the weakly 
nonlinear background. In this way we hope to account properly for the 
weakly nonlinear transition between the linear and strongly nonlinear 
phase of the evolution of the perturbation which lacked in the approach of 
HS. Such a generalization involves constructing the weakly nonlinear 
probability distribution function (PDF) of density around an overdense 
region. The properties of the one-point weakly nonlinear PDFs were 
discussed by many authors (e.g. Bernardeau \& Kofman 1995; Juszkiewicz et 
al. 1995) and all studies confirm that the weakly nonlinear PDFs develop 
features which are absent in the linear phase of the evolution e.g. the 
skewness. Those functions describe the density field at a randomly chosen 
point. Here we impose the condition that the point be chosen in the 
vicinity of a significantly overdense region which requires constructing 
first the two-point weakly nonlinear PDF. Then the PDF in this point is 
given by the conditional probability obtained from the two point PDF with 
the restriction that the density in the second point (the location of the 
peak) is known and equal to a constant. The mean density obtained from 
this weakly nonlinear PDF is then taken as the initial condition for 
spherical collapse. 

The reliability of this approach rests on the assumption that the 
influence of the neighbouring fluctuations can be restricted to the weakly 
nonlinear phase with its only outcome in the form of a changed `initial' 
density profile that can then evolve independently of surroundings, 
according to the spherical model. Although it is quite obvious that the 
evolution of a strongly nonlinear object is governed mainly by its own 
gravity and is little affected by its surroundings, it is very difficult 
to determine precisely the moment when such a situation takes place, i.e. a 
moment when we should pass from the weakly nonlinear statistical 
description to the strongly nonlinear deterministic spherical model.

The paper is organized as follows. In Section~2 we construct the bivariate 
Edgeworth series and calculate the two-point skewness parameter for 
scale-free power-law spectra and Gaussian smoothing. In Section~3 we 
calculate the conditional probability distribution of the density field 
around a peak and discuss its properties. Section~4 provides the 
application of the results to the spherical collapse model. The concluding 
remarks follow in Section~5.

\section{The bivariate Edgeworth series}

We consider density contrast field which initially has Gaussian
distribution with zero mean. The field measured at a given point will be
denoted by the symbol $\delta$ while the one measured at the distance $r$
from the first point will be called $\gamma$.  They can be treated as
two variables, two fields in the same space. Since the fields can in 
general be smoothed with filters of different scales their variances 
can be different: $\langle\delta^{2}\rangle=\sigma^{2}$ and 
$\langle\gamma^{2}\rangle=\tau^{2}$. The (auto)correlation function of 
these two fields is given by
\begin{equation}        \label{l1}
    \langle\delta({\bf x}) \ \gamma({\bf x} + {\bf r})\rangle = \xi(r).
\end{equation}
Hereafter we will consider the normalized density contrast fields $\mu =
\delta / \sigma$ and $\nu = \gamma / \tau$. Their variances
are now equal to unity and their correlation is now given by $\varrho =
\xi / \sigma \tau$. The quantity $\varrho$ will be referred to as the
correlation coefficient. The joint probability distribution of the two
variables in the Gaussian case is
\begin{equation}        \label{l2}
    p(\mu, \nu) = \frac{1}{2 \pi \sqrt{1 - \varrho^2}} \
    f(\mu, \nu, \varrho)
\end{equation}
where
\begin{equation}        \label{l2prim}
    f(\mu, \nu, \varrho) =
    {\rm exp} \left[- \frac{(\mu^{2} - 2 \varrho \mu \nu + \nu^{2})}{2
    (1- \varrho^{2})}  \right].
\end{equation}
If the fields were uncorrelated ($\varrho=0$), $p(\mu, \nu)$ would be just
a product of two Gaussian distributions of $\mu$ and $\nu$.

The purpose of this section is to generalize the joint probability
distribution function in the Gaussian case (\ref{l2}) to the case when the
density fields are weakly nonlinear and therefore departing from
Gaussianity. If the rms fluctuations of the fields are small (below unity)
the fields can be expanded around their linear values $\delta_{1}$ and
$\gamma_{1}$ respectively
\begin{eqnarray}
    \delta & = & \delta_{1} + \delta_{2} + \delta_{3} + \cdots
    \label{l10}\\
    \gamma & = & \gamma_{1} + \gamma_{2} + \gamma_{3} + \cdots.
    \label{l11}
\end{eqnarray}

As discussed by Juszkiewicz et al. (1995) the so-called Edgeworth series 
provides a good approximation to the one-point PDF, except for the very 
tails of the distribution. Up to the third order approximation the 
Edgeworth series reads 
\begin{equation}        \label{l5} 
    p(\nu) = \frac{1}{\sqrt{2 \pi}} \ 
    {\rm e}^{-\nu^{2}/2} \left[ 1 + \frac{1}{3 !} S_{3} \sigma H_{3}(\nu) 
    \right. + \frac{1}{4 !} S_{4} \sigma^{2} H_{4}(\nu) \left. + 
    \frac{10}{6 !} S_{3}^{2} \sigma^{2} H_{6}(\nu) \right] 
\end{equation} 
where $S_{3}$ and $S_{4}$ are respectively the third and fourth normalized
cumulants of the density contrast field, the skewness and the kurtosis
(see e.g. Bernardeau 1994; \L okas et al. 1995) and $H_{n}(\nu)$ is the 
Hermite polynomial of the $n$--th order generated by 
\begin{equation}  \label{l6} 
    (-1)^{n} \frac{{\rm d}^{n}}{{\rm d} \nu^{n}} {\rm 
    e}^{-\nu^{2}/2} = {\rm e}^{-\nu^{2}/2} H_{n}(\nu). 
\end{equation}

In the construction of the bivariate Edgeworth series we follow the
work of Longuet-Higgins (1963, 1964) who considered modified Gaussian
distributions of weakly nonlinear variables and applied them to
statistical theory of sea waves. In the second order approximation we
have
\begin{equation}
    p(\mu, \nu) = \frac{1}{2 \pi \sqrt{1 - \varrho^{2}}} \
    f(\mu, \nu, \varrho)
    \left[ 1 + \frac{1}{6} (\lambda_{30} H_{30} + 3 \lambda_{21}
    H_{21} + 3 \lambda_{12} H_{12} + \lambda_{03} H_{30}) \right]
    \label{l7}
\end{equation}
where $f(\mu, \nu, \varrho)$ is given by equation (\ref{l2prim}).
The correlation coefficient $\varrho$, the Hermite polynomials $H_{mn}$ 
and the two-point moments $\lambda_{mn}$ will be discussed in the 
following.

\subsection{The correlation coefficient}

The autocorrelation function of density field measured at two points
separated by distance $|{\bf r}| = r$ can be calculated using the
relation
\begin{equation}    \label{l22prim}
    \xi_{R}(r) = \int \frac{{\rm d}^{3} k}{(2 \pi)^{3}}
    P(k) W^{2}(k R) \ {\rm e}^{i {\bf k \cdot r}}
\end{equation}
where $P(k)$ is the power spectrum of the density fluctuations and the
field is smoothed with a window function $W$ of radius $R$ which in 
what follows will be assumed to have a Gaussian form 
\begin{equation}  \label{e12}
    W(kR)={\rm e}^{-k^{2} R^{2}/2}.
\end{equation}
In what follows we will restrict the calculations to the case when the 
field at both locations has the same properties and is smoothed with the 
same smoothing radius so that the variances are equal: $\sigma = \tau$.

\begin{figure}[p]  
\begin{center}
    \leavevmode
    \epsfxsize=2.8in
    \epsfbox[155 15 440 785]{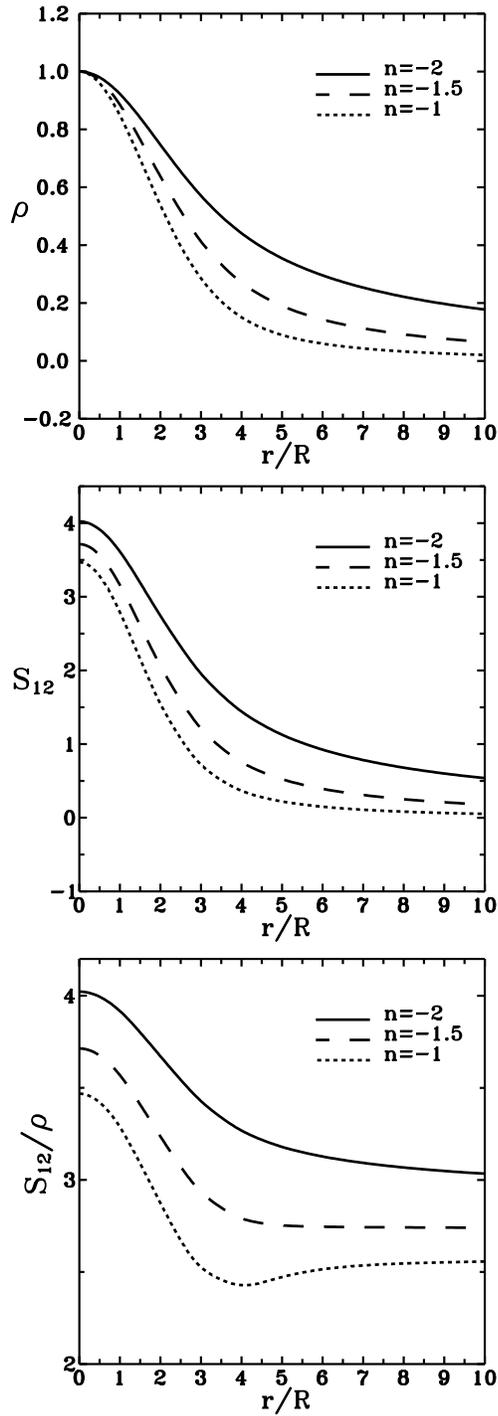}
\end{center}
    \caption{The values of the correlation coefficient $\varrho$, the 
    two-point skewness $S_{12}$ and the ratio $S_{12}/\varrho$ as 
    functions of the distance between the two points (in units of the 
    smoothing scale), $c = r/R$, for different power spectra.} 
\label{s12ro} 
\end{figure}

For the scale-free power spectra 
\begin{equation}  \label{e15}
    P(k) = C k^n, \ \ \ -3 \le n \le 1
\end{equation}
we obtain
\begin{equation}    \label{l24prim}
    \varrho = \frac{\xi_{R}(r)}{\sigma^2} = \frac{\sqrt{\pi}}{2}
    \ \frac{\Gamma(1 + \frac{n}{2})}{\Gamma(\frac{n+3}{2})} \
    L_{n/2}^{1/2} \left( \frac{c^2}{4} \right) \ {\rm e}^{-c^2/4}
\end{equation}
where $c=r/R$ and $L_{\beta}^{\alpha}(x)$ are Laguerre polynomials.
Here we have used the fact that for the scale-free power spectra 
(\ref{e15}) and Gaussian smoothing (\ref{e12}) the linear variance of the 
density PDF is given by 
\begin{equation}    \label{e20} 
  \sigma^{2} = \langle\delta^{2}_{1}\rangle = 
  D^{2}(t) \int \frac{{\rm d}^3 k}{(2 \pi)^{3}} P(k) W^{2}(k R) =
  C D^{2}(t) \frac{\Gamma(\frac{n+3}{2})}{(2 \pi)^{2} R^{n+3}}. 
\end{equation}
Hereafter the variance $\langle\delta^{2}\rangle$ will be always 
approximated by the linear value (\ref{e20}) which is sufficient in the 
construction of the two-point Edgeworth series up to second order. The 
result (\ref{l24prim}) takes simpler form in terms of the degenerate 
hypergeometric function
\begin{equation}    \label{l26prim}
    \varrho(n, c) = \ _{1}F_{1} \left( \frac{n+3}{2}, \frac{3}{2},
    - \frac{c^2}{4} \right).
\end{equation}
We see that the correlation coefficient depends on the scales $r$ and $R$
only via their ratio $c$. Upper panel of Figure~\ref{s12ro} shows the shape
of $\varrho(c)$ for different power spectrum indices $n=-2,-1.5,-1$.

\subsection{The bivariate Hermite polynomials}

The quantities $H_{mn}$ in (\ref{l7}) are two-dimensional analogues of the 
Hermite polynomials $H_{n}$ appearing in the one-point Edgeworth series 
(\ref{l5}) that were generated by equation (\ref{l6}). The bivariate 
Hermite polynomials can be calculated using similar formula
\begin{equation}
    (-1)^{m+n} \frac{{\rm d}^{m}}{{\rm d} \mu^{m}}
    \frac{{\rm d}^{m}}{{\rm d} \nu^{n}} f(\mu, \nu, \varrho)
    = f(\mu, \nu, \varrho) H_{mn}(\mu, \nu, \varrho).
    \label{l16}
\end{equation}
The polynomials of orders needed in this calculation can be
written in the form
\begin{equation}        \label{l17}
    H_{mn} (\mu, \nu, \varrho) = \frac{1}{(1 - \varrho^{2})^{2}}
    h_{mn} (\mu, \nu, \varrho)
\end{equation}
where
\begin{eqnarray}
    h_{30} & = & \frac{(\mu - \varrho \nu)^{3}}{1 - \varrho^{2}}
    - 3 (\mu - \varrho \nu) \nonumber\\
    h_{21} & = & 2 \varrho (\mu - \varrho \nu) - (\nu - \varrho \mu)
    + \frac{(\nu - \varrho \mu) (\mu - \varrho \nu)^{2}}{1 - \varrho^{2}}
    \label{l18}\\
    h_{12} & = & 2 \varrho (\nu - \varrho \mu) - (\mu - \varrho \nu)
    + \frac{(\mu - \varrho \nu) (\nu - \varrho \mu)^{2}}{1 - \varrho^{2}}
    \nonumber\\
    h_{03} & = & \frac{(\nu - \varrho \mu)^{3}}{1 - \varrho^{2}}
    - 3 (\nu - \varrho \mu). \nonumber
\end{eqnarray}

\subsection{The two-point skewness}

The coefficients $\lambda_{ij}$ in (\ref{l7}) are defined by the reduced
moments (cumulants) $\kappa_{ij}$ in the following way
\begin{equation}        \label{l8}
    \lambda_{ij} = \frac{\kappa_{ij}}{(\kappa_{20}^{i}
    \kappa_{02}^{j})^{1/2}}
\end{equation}
while the cumulants of the fields $\delta$ and $\gamma$ are given by
the connected part of the moments
\begin{equation}        \label{l9}
    \kappa_{ij} = \langle\delta^{i} \gamma^{j}\rangle_{c}.
\end{equation}

Using the perturbative expansion of the fields, equations
(\ref{l10})-(\ref{l11}), we find up to second order
\begin{eqnarray}
    \kappa_{30} & = & \langle\delta_{1}^{3}\rangle + \ 3 
    \langle\delta_{1}^{2} \delta_{2}\rangle \nonumber\\ \kappa_{21} & = & 
    \langle\delta_{1}^{2} \gamma_{2}\rangle + \ 2 \langle\delta_{1} 
    \delta_{2} \gamma_{1}\rangle \label{l12}\\ 
    \kappa_{12} & = & \langle\delta_{2} \gamma_{1}^{2}\rangle + \ 2 
    \langle\delta_{1} \gamma_{1} \gamma_{2}\rangle \nonumber\\ 
    \kappa_{03} & = & \langle\gamma_{1}^{3}\rangle + \ 3 
    \langle\gamma_{1}^{2} \gamma_{2}\rangle. \nonumber 
\end{eqnarray} 
The moments $\kappa_{20}$ and $\kappa_{02}$ are just the variances 
$\langle\delta^{2}\rangle$ and $\langle\gamma^{2}\rangle$ 
respectively, which, as stated before, can be both approximated by the 
linear value $\sigma^{2}$ given by equation (\ref{e20}).

The moments $\kappa_{30}$ and $\kappa_{03}$ are related to the ordinary
one-point skewness parameters (see e.g. Peebles 1980; Juszkiewicz et al. 
1993). For Gaussian fields $\langle\delta_{1}^{3}\rangle = 0$, and we have
\begin{equation}    \label{l13}
  \kappa_{30} = 3 \langle\delta_{1}^{2} \delta_{2}\rangle = S_{3} 
  \sigma^{4}
\end{equation}
where $S_{3}$ is the dimensionless skewness parameter which
for a Gaussian filter depends on the spectral index $n$ in the following 
way (\L okas et al. 1995)
\begin{equation}     \label{e26}
   S_{3} = 3
   \ \ _{2}F_{1} \left( \frac{n+3}{2},\frac{n+3}{2}, \frac{3}{2}, \right.
   \left. \frac{1}{4} \right) - \left( n +\frac{8}{7} \right) \
   _{2}F_{1} \left( \frac{n+3}{2},\frac{n+3}{2}, \frac{5}{2}, \right.
   \left. \frac{1}{4} \right).
\end{equation}
The other one-point moment $\kappa_{03} = S_{3} \tau^{4}$ but since we 
have already put $\tau = \sigma$ we have $\kappa_{03} = \kappa_{30}$.

The two-point moments $\kappa_{21}$ and $\kappa_{12}$ can be calculated
using the same method as in the case of one-point skewness (\L okas et 
al. 1995) but now we have to keep the dependence of the moments on 
distance $r = |{\bf r}|$. The two-point moments also scale as $\sigma^{4}$ 
so it is convenient to calculate them in the normalized form. Using 
equation~(\ref{e20}) we have for $\kappa_{21}/\sigma^{4}$ 
\begin{eqnarray}
    \frac{\langle\delta_{1}^{2} \gamma_{2}\rangle(r)}{\sigma^4} & = & 
    \frac{1}{448 \pi^6 R^{2(n+3)} \sigma^{4}} 
    \int {\rm d}^3 p \int {\rm d}^3 q \ W(p) W(q) W(|{\bf p+q}|) 
    {\rm e}^{{\rm i}\: ({\bf p+q}) \cdot {\bf r}} \times \label{l14}\\ 
    & \times & P(p) P(q) J({\bf p+q,p,q}) \nonumber 
\end{eqnarray} 
and 
\begin{eqnarray}
    \frac{2 \langle\delta_{1} \delta_{2} \gamma_{1}\rangle(r)}{\sigma^4} 
    & = & \frac{1}{224 \pi^6 R^{2(n+3)} \sigma^{4}} \int {\rm d}^3 p \int 
    {\rm d}^3 q \ W(p) W(q) W(|{\bf p+q}|) {\rm e}^{{\rm i}\: {\bf p} 
    \cdot {\bf r}} \times \label{l15} \\ 
    & \times & P(p) P(q) J({\bf p+q,p,q})  \nonumber
\end{eqnarray}
where $J$ is the kernel of the second order perturbative solution for 
density field. The integrals (\ref{l14}) and (\ref{l15}) will be evaluated 
for scale-free power spectra (\ref{e15}) and Gaussian smoothing 
(\ref{e12}). The simplest way to perform the integrals in (\ref{l14}) is 
to change variables from {\bf p}, {\bf q} to {\bf l}, {\bf p} so that {\bf 
p} + {\bf q} = {\bf l} and $ {\bf l \cdot p} = l\: p \: \beta$. Together 
with the expression for the variance (\ref{e20}) this produces
\begin{eqnarray}
    \frac{\langle\delta_{1}^{2} \gamma_{2}\rangle(c)}{\sigma^4} 
    & = & \frac{2}{7 \ \Gamma^2(\frac{n+3}{2}) \ c} \int {\rm d} p \int 
    {\rm d} l \ {\rm sin}(c l) \ p^{n+2} \ l \ {\rm e}^{-l^2 - p^2} \times
    \label{l14prim} \\
     & \times & \int_{-1}^{1} {\rm d} \beta \ J(l,p,\beta)
    \ (l^2 + p^2 - 2 l p \beta)^{n/2} \ {\rm e}^{l p \beta} \nonumber
\end{eqnarray}
where $c = r/R$ as before. The integration over the angular variable
$\beta$ can still be done analytically for integer or half-integer values
of the spectral index $n$. The remaining integrations are done
numerically. In the case of (\ref{l15}) the integration over angular
variables gives 
\begin{eqnarray}
    \frac{2 \langle\delta_{1} \delta_{2} \gamma_{1}\rangle(c)}{\sigma^4} 
    & = &
    \frac{8 \sqrt{\pi}}{\sqrt{2} \ \Gamma^{2}(\frac{n+3}{2}) \ c} \
    \int {\rm d} p \int {\rm d} q \ {\rm sin}(c p) \ p^{n+1/2} \ q^{n+3/2}
    \ {\rm e}^{-p^2 - q^2} \times \label{l15prim} \\
    & \times & \left[\frac{34}{21} I_{1/2}(p q) - \left(\frac{p}{q} +
    \frac{q}{p} \right) I_{3/2}(p q) + \frac{8}{21} I_{5/2}(p q) \right].
    \nonumber
\end{eqnarray}
By using the expansion of Bessel functions 
\begin{equation}      \label{e25}
   I_{\nu}(z) = \sum_{m=0}^{\infty} \frac{1}{m! \Gamma(\nu + m+1)}
   \left( \frac{z}{2} \right)^{\nu+2 m} 
\end{equation}
and the facts that
\begin{eqnarray}
    \int_{0}^{\infty} q^{a} {\rm e}^{-q^2} \ {\rm d} q & = &
    \frac{1}{2} \ \Gamma \left(\frac{a+1}{2} \right) \ \
    {\rm for} \ \ a>-1 \\      \label{l15aa}
    \int_{0}^{\infty} p^{b} {\rm e}^{-p^2} \ {\rm sin} (c p) \ {\rm d} p
    & = & \frac{c}{2} \ \Gamma \left( 1+\frac{b}{2} \right) \ \ _{1} F_{1}
    \left( 1 + \frac{b}{2}; \frac{3}{2}; - \frac{c^2}{4} \right) \ \
    {\rm for} \ \ b>-2
    \label{l15bb}
\end{eqnarray}
we obtain the result in the form of a series of combination of gamma
and degenerate hypergeometric functions or Laguerre polynomials
\begin{eqnarray}
    \frac{2 \langle\delta_{1} \delta_{2} \gamma_{1}\rangle(c)}{\sigma^4} 
    & = & \frac{\pi}{2 \Gamma^{2}(\frac{n+3}{2})}\ {\rm e}^{-c^2/4} 
    \sum_{m=0}^{\infty} \frac{\Gamma (\frac{n+3}{2}+m)}{m! \ 2^{2m}}
    \times \nonumber \\
    & \times &
    \left\{ \frac{34}{21} \frac{\Gamma(\frac{n+2}{2}+m)}{\Gamma
    (m+\frac{3}{2})}  \right.
    \ L_{n/2+m}^{1/2} \left( \frac{c^2}{4} \right)
    \nonumber \\
    & - & \frac{1}{2 \Gamma(m+\frac{5}{2})}
    \left[ \Gamma \left( \frac{n+4}{2}+m \right) \right.
    \ L_{n/2+m+1}^{1/2} \left( \frac{c^2}{4} \right)
    \label{l15c} \\
    & + & \left( \frac{n+3}{2}+m \right) \Gamma
    \left( \frac{n+2}{2}+m \right)
    \left. \ L_{n/2+m}^{1/2} \left( \frac{c^2}{4} \right) \right]
    \nonumber \\
    & + & \frac{8}{21} \frac{(\frac{n+3}{2}+m) \Gamma(\frac{n+4}{2}+m)}
    {4 \Gamma(m+\frac{7}{2})}
    \left. \ L_{n/2+m+1}^{1/2} \left( \frac{c^2}{4} \right) \right\}.
    \nonumber
\end{eqnarray}
The values of
\begin{equation}   \label{l15bis}
    S_{21} = \frac{\kappa_{21}}{\sigma^{4}}
\end{equation}
will be hereafter called the two-point skewness parameter. Because of the
symmetry between the two locations we obviously have $S_{12} = S_{21}$. In
the following calculations we will therefore use only the symbol $S_{12}$.
The numerical values of $S_{12}$ for different values of $c$ and
$n=-2,-1.5,-1$ are given in Table~\ref{s12} and plotted in the middle 
panel of Figure~\ref{s12ro}.

\begin{table}
\centering
\begin{tabular}{llll}  \hline \hline
  & &   $S_{12}$   & \\
  $c=r/R$ & $n=-2$ & $n=-1.5$ & $n=-1$ \\
  \hline
\ 0    &  4.022  & 3.714    & 3.468     \\
\ 0.01 &  4.022  & 3.714    & 3.468     \\
\ 0.1  &  4.018  & 3.708    & 3.460     \\
\ 0.5  &  3.913  & 3.563    & 3.282     \\
\ 1  &    3.614  & 3.159    & 2.790     \\
\ 2  &    2.742  & 2.062    & 1.547     \\
\ 3  &    1.957  & 1.212    & 0.7212    \\
\ 4  &    1.441  & 0.7545   & 0.3659    \\
\ 5  &    1.126  & 0.5226   & 0.2206    \\
\ 6  &    0.9237 & 0.3914   & 0.1494    \\
\ 7  &    0.7830 & 0.3081   & 0.1084    \\
\ 8  &    0.6796 & 0.2510   & 0.08233   \\
\ 9  &    0.6003 & 0.2097   & 0.06471   \\
  10 &    0.5377 & 0.1787   & 0.05222   \\
\hline
\hline
\end{tabular}
\caption{The values of the two-point skewness parameter $S_{12}$ for 
density field as a function of the spectral index $n$ and $c=r/R$.}
\label{s12}
\end{table}

The lower panel of Figure~\ref{s12ro} shows the ratio $S_{12}/\varrho$.
As discussed by Bernardeau (1996) for the case of top-hat filter such
quantity should approach a constant at $c \rightarrow \infty$. 
Figure~\ref{s12ro} shows that the same behaviour is observed for Gaussian 
filtering. Using diagrammatic representations of the two-point moments 
Bernardeau (1996) has shown that the part of $S_{12}$ given by equation 
(\ref{l14}) behaves like $\varrho^{2}$ in this limit and therefore is 
negligible compared to the part given by equation (\ref{l15}) which 
behaves like $\varrho$. For large $c$ the value of the sum in (\ref{l15c}) 
is perfectly approximated by the $m=0$ element of the series and we find 
\begin{equation}  \label{l15a}
   S_{12} = s_{12} \varrho
\end{equation}
where
\begin{equation}  \label{l15b}
   s_{12} = \frac{47}{21} - \frac{n}{3}.
\end{equation}
It is worth noting that the moment $s_{12}$ given by equation (\ref{l15b})
is exactly equal to the corresponding quantity $C_{2,1}$ calculated by
Bernardeau (1996) for scale-free power spectra and top-hat smoothing.

For $n=-2, -1.5, -1$ the numerical values of $s_{12}(n)$ are respectively:
2.90, 2.74 and 2.57. It is seen in Figure~\ref{s12ro} that the values of
$S_{12}/\varrho$ indeed approach them for large $c$. The value of $S_{12}$
is well approximated by the one given by equation (\ref{l15a}) within 10\%
for $c > 4$ (and the approximation works much better for $n=-1$ and
$n=-1.5$ than for $n=-2$). In the following, however, we will be also
interested in the region of $c$ of the order of few, where this
approximation is not sufficient and cannot be adopted.

In the limit of $c \rightarrow 0$, which is equivalent to $r
\rightarrow 0$, the two points at which we measure the density
field converge and we expect that the two-point skewness reproduces the
one-point value of $S_{3}$ given by equation (\ref{e26}). As Table~1
and Figure~\ref{s12ro} prove this is indeed the case. In this limit the 
two parts (\ref{l14}) and (\ref{l15}) have comparable contributions: 
$S_{3}/3$ and $2 S_{3}/3$ respectively.

Using the results of this section the bivariate Edgeworth series
(\ref{l7}) can be rewritten in the following way
\begin{eqnarray}
    p(\mu, \nu) &=& \frac{1}{2 \pi \sqrt{1 - \varrho^{2}}} \
    f(\mu, \nu, \varrho)     \label{l19}\\
    & \times & \left\{ 1 + \frac{\sigma}{6 (1-\varrho^{2})^{2}}
    \left[ S_{3} (h_{30} + h_{03}) + 3 S_{12} (h_{12} + h_{21})
    \right] \right\}. \nonumber
\end{eqnarray}

\section{Conditional probability distribution around a peak}

Let us now suppose that the value of one variable is known: $\nu = a$.
The conditional probability distribution of the other variable is defined
as
\begin{equation}        \label{l3}
    p(\mu, \nu=a) = \frac{p(\mu, a)}{\int_{- \infty}^{+\infty} \
    p(\mu, a) \ {\rm d} \mu}
\end{equation}
where the nominator is the bivariate distribution with the second
variable put equal to a constant and the denominator is the marginal
distribution of the variable $\nu = a$.

In the case of two-point Gaussian distribution (\ref{l2}) the resulting
conditional distribution for $\mu$ reads 
\begin{equation}        \label{l4}
    p(\mu, \nu=a) = \frac{1}{\sqrt{2 \pi (1 - \varrho^{2})}}
    \ {\rm exp} \left[- \frac{(\mu - \varrho a)^{2}}{2 (1- \varrho^{2})}
    \right].
\end{equation}
This is again a Gaussian but the variance was changed from unity to
$1 - \varrho^{2}$ and the mean value of $\mu$ was moved from zero
to $\mu = \varrho a$.

In the case of two-point Edgeworth series (\ref{l19}) the conditional
distribution of $\mu$, provided the value of $\nu$ is known and equal to
$a$, is obtained from equation (\ref{l3}) using the marginal distribution
of $\nu$
\begin{equation}        \label{l20}
    \int_{- \infty}^{+\infty} \ p(\mu, \nu=a) \ {\rm d} \mu =
    \frac{1}{\sqrt{2 \pi}} {\rm e}^{-a^{2}/2}
    \left[ 1 + \frac{S_{3} \sigma}{6} (a^3 - 3 a) \right]
\end{equation}
which is just the one-point Edgeworth series (\ref{l5}) up to
second order with $\nu=a$. Since only the lowest order correction in the
Edgeworth series was introduced the result must be expanded in powers of
$\sigma$ and only the term linear in $\sigma$ should be kept
\begin{eqnarray}       \label{l21}
    p(\mu, \nu=a) &=& \frac{1}{\sqrt{2 \pi (1 - \varrho^{2})}}
    \ {\rm exp} \left[- \frac{(\mu - \varrho a)^{2}}{2 (1- \varrho^{2})}
    \right]  \times \\
    & \times & \left\{ 1 + \frac{\sigma}{6 (1-\varrho^{2})^{2}}
    \left[ S_{3} (h_{30} + h_{03}) + 3 S_{12} (h_{12} + h_{21}) \right]
    - \frac{S_{3} \sigma}{6} (a^{3} - 3 a) \right\} \nonumber
\end{eqnarray}
where $h_{mn} = h_{mn}(\mu, a, \varrho)$ are given by equations
(\ref{l18}).

First let us check the behaviour of the conditional probability 
(\ref{l21}) in the limit of the correlation coefficient $\varrho 
\rightarrow 0$ i.e. when the two considered points are separated by large 
distance and therefore uncorrelated. According to the results of the 
previous section in this limit also $S_{12} \rightarrow 0$ and the 
distribution (\ref{l21}) becomes
\begin{equation}        \label{l22}
    p(\mu, \nu=a) |_{\varrho \rightarrow 0} =
    \frac{1}{\sqrt{2 \pi}} \ {\rm e}^{-\mu^{2}/2} \
    \left[ 1+\frac{S_{3} \sigma}{6} (\mu^{3} - 3 \mu) \right]
\end{equation}
which is just the one-point Edgeworth series (\ref{l5}), as expected.
The probability distribution for the density field is not supposed to
depend on a particular value of density contrast at infinite distance.

In order to test the behaviour of the distribution (\ref{l21}) for
different parameters $\varrho$, $a$ and $\sigma$ we calculate its lowest
order moments with respect to the mean value $\langle\mu\rangle_{E}$ (the
subscript $E$ will identify the quantities obtained for the distribution
(\ref{l21})) and compare them with the corresponding quantities of the
Gaussian conditional distribution (\ref{l4}). Let us recall that for
Gaussian distribution we found (in agreement with Dekel 1981, Peebles 1984
and HS)
\begin{eqnarray}
    \langle\mu\rangle_{G} &=& \varrho \ a                    
    \label{l24}\\
    \langle(\mu - \langle\mu\rangle)^{2}\rangle_{G} &=& 1 - \varrho^{2}  
    \label{l25}\\
    \langle(\mu - \langle\mu\rangle)^{3}\rangle_{G} &=& 0.               
    \label{l26}
\end{eqnarray}
After tedious but straightforward calculations for the Edgeworth
conditional probability (\ref{l21}) we get
\begin{eqnarray}
    \langle\mu\rangle_{E} &=& \varrho \ a + \frac{\sigma}{2} (a^{2} - 1)
    (S_{12} - \varrho S_{3})
    \label{l27}\\
    \langle(\mu - \langle\mu\rangle)^{2}\rangle_{E} &=& 1 - \varrho^{2} +
    \sigma \ a \ (S_{12} - 2 \varrho S_{12} + \varrho^{2} S_{3})
    \label{l28}\\
    \langle(\mu - \langle\mu\rangle)^{3}\rangle_{E} &=& \sigma \ 
    (S_{3} - 3 \varrho S_{12}
    + 3 \varrho^{2} S_{12} - \varrho^{3} S_{3}).
    \label{l29}
\end{eqnarray}
From equation ({\ref{l28}) we also have the dispersion to the lowest order
in $\sigma$
\begin{equation}   \label{l30}
    \langle(\mu - \langle\mu\rangle)^{2}\rangle_{E}^{1/2} = 
    \sqrt{1 - \varrho^{2}} + \frac{\sigma \ a \ (S_{12} - 2 \varrho 
    S_{12} + \varrho^{2} S_{3})} {2 \sqrt{1- \varrho^{2}} }. 
\end{equation}

Clearly the characteristics of the conditional Edgeworth distribution are
given by the Gaussian ones plus correction terms proportional to $\sigma$
and a function of $a$ (except for the third moment which is independent of
$a$). As shown earlier in equation (\ref{l22}) in the limit of $\varrho 
\rightarrow 0$ the conditional Edgeworth distribution approaches one-point 
Edgeworth series. This must also apply to its moments. Using the fact that 
at small $\varrho$ the values of two-point skewness $S_{12}$ vanish we 
again find that in this limit the third order moment (\ref{l29}) 
approaches the value of the one-point Edgeworth series, $S_{3} \sigma$. 
The same is true for the lower moments (\ref{l27}) and (\ref{l28}).

An independent check of the results (\ref{l27})-(\ref{l29}) is provided in
the limit of $\varrho \rightarrow 1$. Then the distribution $p(\mu,
\nu=a)$ should approach the Dirac's delta function, $\delta_{D}(\mu - a)$,
with the mean equal to $a$ and the second and third moment equal to zero.
Recalling that in this limit $S_{12} \rightarrow S_{3}$ such results can
be immediately reproduced from equations (\ref{l27})-(\ref{l29}).

An interesting application of these results from the point of view
of the theory of structure formation is to study the density distribution
around an overdense region. Although the results presented so far apply to
arbitrary values of $a$ we will from now on focus on regions which are
able to dominate their surroundings, that is their density contrast is 
bigger than one standard deviation. This corresponds to assuming $a > 1$. 
A region of such overdensity will be called a peak although it might not 
be a maximum in a strict mathematical sense. However, we expect that 
points of $a \gg 1$ most probably correspond to local maxima. As reasoned 
by HS, this is also true for mild values of $a \simgt 1$ (see also Adler 
1981).

If $a > 1$ the moments (\ref{l24})-(\ref{l29}) can be interpreted as
characteristics of the density distribution around a density peak. The
Gaussian values are the well known results of linear theory while the
values for the Edgeworth conditional distribution provide the corrections
introduced by the fact that the field on a given smoothing scale has
become weakly nonlinear. The direction of the effect depends on the
numerical values of the one-point and two-point skewness parameters and
the correlation coefficient.

\begin{figure}[p]  
\begin{center}
    \leavevmode
    \epsfxsize=2.8in
    \epsfbox[155 15 440 785]{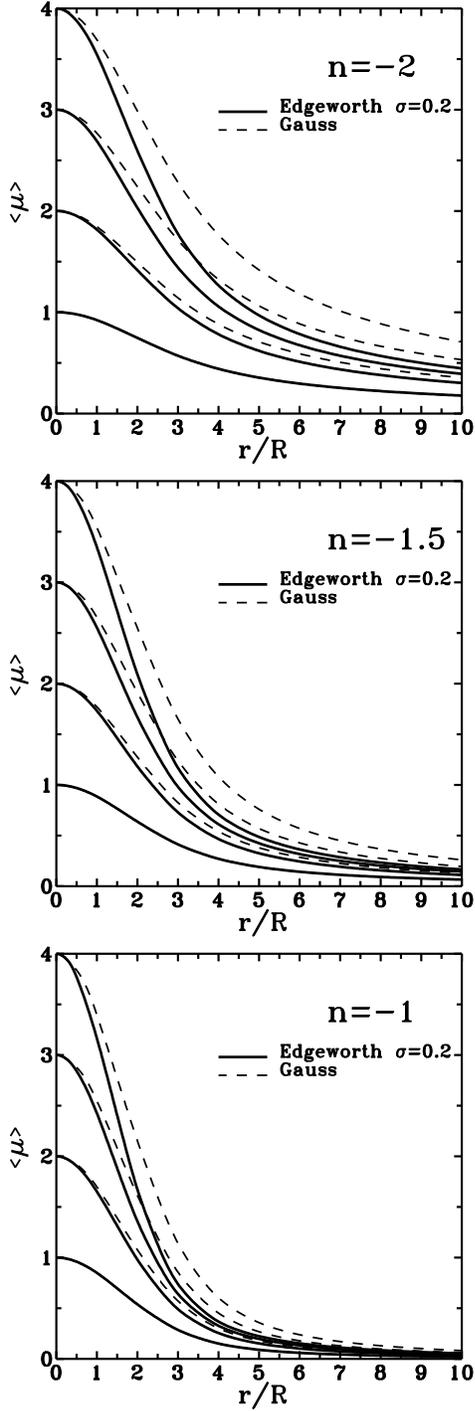}
\end{center}
    \caption{The mean normalized density contrast as a function of $c = 
    r/R$. The dashed lines correspond to the Gaussian case while the solid 
    ones show the results obtained from the Edgeworth approximation with 
    $\sigma = 0.2$. Each panel shows results for $a=1,2,3,4$ and 
    different power spectrum.} 
\label{s} 
\end{figure}

It is clear from Figure~\ref{s12ro} that for the spectral indices 
considered here ($n=-2,-1.5,-1$) and nonzero distance from 
the peak we have $S_{12}/\varrho < S_{3}$ which can be rewritten as
\begin{equation}  \label{l30a}
    S_{12} - \varrho S_{3} < 0.
\end{equation}
This proves that according to equation (\ref{l27}) for $a>1$ the
correction to the mean normalized density with respect to the Gaussian
value is always negative. Figure~\ref{s} shows the expected normalized 
density contrast $\langle\mu\rangle$ for the Edgeworth (\ref{l27}) and 
Gaussian (\ref{l24}) conditional distributions as a function of the 
distance from the peak (in units of the smoothing scale), $c=r/R$. Each 
panel shows results for different scale-free power spectrum with spectral 
index $n = -2,-1.5,-1$.  In each panel the Edgeworth results are plotted 
as thicker solid lines while the thinner dashed lines show the Gaussian 
ones. The four pairs of lines in each panel correspond from bottom to top 
to the four chosen values of the peak's height: $a=1,2,3,4$ (for $a=1$ the 
results are the same in both Gaussian and Edgeworth cases). The Edgeworth 
results are always below the Gaussian values so it is clear that the 
effect of weakly nonlinear interactions is to decrease the expected density 
around an overdense region. Equation (\ref{l27}) shows explicitly that the 
effect grows with $a$ and $\sigma$.

Juszkiewicz et al. (1995) have tested the third order Edgeworth 
approximation against N-body simulations and found that it is accurate for 
density contrast up to $\delta = a \sigma = 1$. Here we are using only the 
second order approximation so the range of validity of these results is 
probably even more restricted. Therefore in the following it will be 
assumed that $a \sigma$ is always below unity. All the Edgeworth results 
in Figure~\ref{s} are given for $\sigma=0.2$. At this value of the rms 
density fluctuation we still have $a \sigma < 1$ even for $a=4$ so we can 
expect the Edgeworth series to be a good approximation. 

\begin{figure}[p]  
\begin{center}
    \leavevmode
    \epsfxsize=2.8in
    \epsfbox[155 15 440 785]{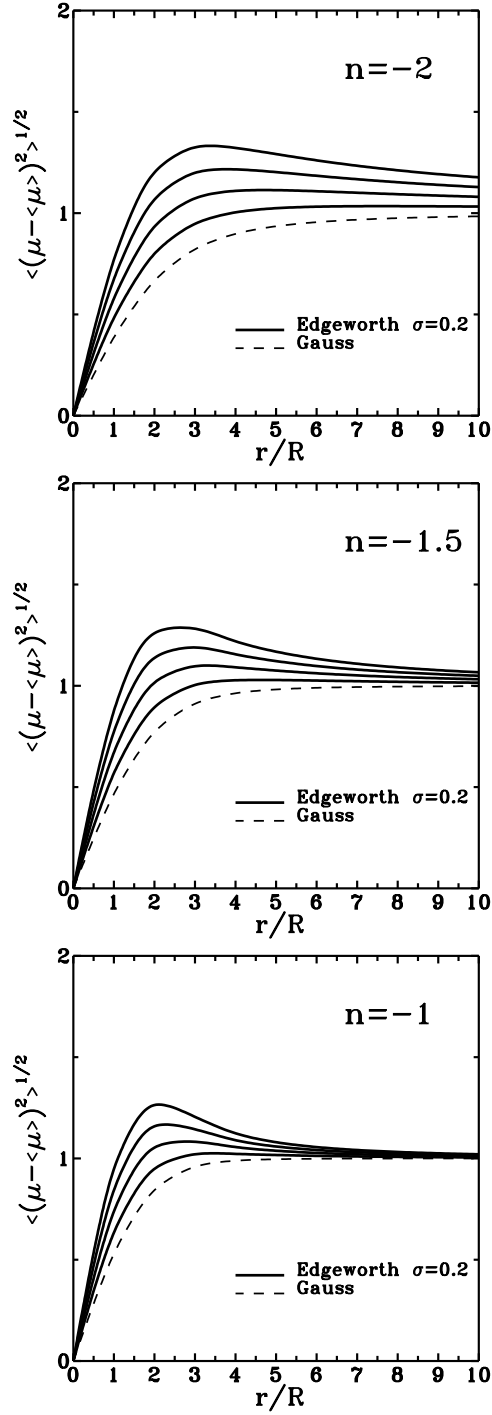}
\end{center}
    \caption{The dispersion of the normalized density contrast as a 
    function of the distance from the peak, $c = r/R$. The dashed line in 
    each panel shows Gaussian results which are independent of the peak's 
    height while the four solid lines correspond to the results obtained 
    from the Edgeworth approximation with $\sigma=0.2$ and $a=1,2,3,4$. 
    Each panel shows results for different power spectrum. } 
\label{v} 
\end{figure}

In the case of the variance the weakly nonlinear corrections work in the 
opposite direction: their effect is to increase the value of the variance 
(or dispersion) with respect to the linear case, because for the power 
spectra considered here (and nonzero distances from the peak) we always 
have
\begin{equation}  \label{l30b}
    S_{12} - 2 \varrho S_{12} + \varrho^{2} S_{3} =
    (1-\varrho)S_{12} + \varrho (\varrho S_{3} - S_{12}) > 0
\end{equation}
where the inequality (\ref{l30a}) was used. The effect grows linearly with
$\sigma$ and $a$ as equation (\ref{l28}) states. In Figure~\ref{v} we plot 
the dispersion of the conditional Edgeworth distribution (\ref{l30}) and 
the Gaussian dispersion which is independent of $a$ and equal to $\sqrt{1 
- \varrho^{2}}$. The quantities are shown as functions of $c=r/R$ for 
$\sigma=0.2$, $a=1,2,3,4$ and different power spectra. As in 
Figure~\ref{s} the thicker solid lines in each panel correspond to the 
Edgeworth results and the thinner dashed lines to the Gaussian ones. 
Although the Gaussian dispersion is always below unity and approaches it 
at large $c$, the Edgeworth values reach a maximum, which is above unity 
at $c$ of the order of a few and then decrease down to unity at large 
distances.

An important conclusion coming from the behaviour of weakly nonlinear 
expected density and its variance is the following. It is clear that a 
local density peak that rises significantly above the noise should 
gravitationally dominate its surroundings out to some distance. A
reasonable measure of the distance, up to which a coherent structure
around the peak is expected, is the scale $r_{coh}$ at which
\begin{equation}  \label{l31}
    \langle\mu\rangle = \langle(\mu - \langle\mu\rangle)^{2}\rangle^{1/2}.
\end{equation}
Because of smoothing it is more convenient to measure the distance $r$ in 
units of the smoothing scale $R$ so in what follows we will use 
$c_{coh} = r_{coh}/R$ instead of $r_{coh}$. The mean 
$\langle\mu\rangle$ as a function of $c$ up to $c_{coh}$ may be 
therefore treated as the density profile of matter that is bound to the 
peak in a statistical sense.

\begin{figure}[p]  
\begin{center}
    \leavevmode
    \epsfxsize=2.8in
    \epsfbox[155 15 440 785]{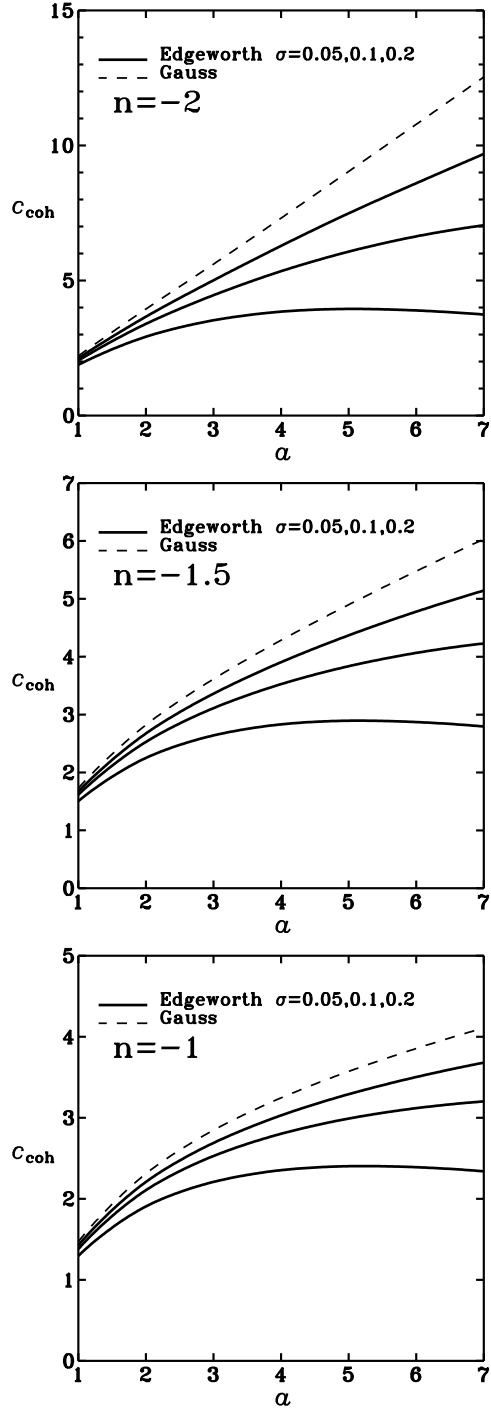}
\end{center}
    \caption{The length of coherence (in units of the smoothing scale) 
    defined by equation~(\protect\ref{l31}) 
    as a function of the peak's height, 
    $a$. The dashed line in each panel shows the results in the Gaussian 
    case. The solid lines correspond to the results obtained from the 
    Edgeworth approximation with $\sigma = 0.05,0.1,0.2$, with larger 
    $\sigma$ producing a curve that departs more from the Gaussian one.} 
\label{ccoh} 
\end{figure}

In the Gaussian case equation (\ref{l31}) leads to 
$\varrho(n, c_{coh})= 1/\sqrt{a^{2} + 1}$ which can be solved 
numerically for $c_{coh}$ given the shape of $\varrho$, equation 
(\ref{l26prim}). Finding $c_{coh}$ in the case of Edgeworth approximation 
requires fitting the function $S_{12}(c)$ and then equation (\ref{l31}) 
can also be solved numerically. The results for both cases are shown in 
Figure~\ref{ccoh} as dependent on the height of the peak $a$, for three 
different values of $\sigma=0.05,0.1,0.2$ and three spectral indices 
$n=-2,-1.5,-1$. Since the mean density is decreased and the dispersion is 
increased the effect of nonlinearity is always to decrease the coherence 
length. The magnitude of the effect grows with the rms fluctuation 
$\sigma$. Therefore the density field around a peak embedded in the weakly 
nonlinear field decorrelates much faster with distance than in the linear 
case.

Note that the scale of each panel in Figure~\ref{ccoh} is different: the 
values of $c_{coh}$ are generally significantly larger for lower spectral 
indices but this reflects mainly the amount of large-scale power in the 
initial linear power spectrum. To take this into account the weakly 
nonlinear coherence length should be compared to the corresponding 
Gaussian one. Then the dependence on the power spectrum is clear: for a 
given $a$ and $\sigma$ the change in the value of $c_{coh}$ with respect 
to the Gaussian one is larger for lower spectral indices. For example for 
$a=3$ and $\sigma=0.1$ the weakly nonlinear $c_{coh}$ is decreased with 
respect to the linear value by 11\%, 14\% and 21\% for $n=-1, -1.5, -2$ 
respectively.

In general the coherence length grows with the height of the peak, $a$,
which is reasonable: higher peaks should dominate their surroundings to
farther extent. However, for largest values of $\sigma=0.2$ we observe
that the curves in Figure~\ref{ccoh} have a maximum at $a \approx 5$. 
Since in this case $a \sigma = 1$ it is probably the effect of breakdown 
of the Edgeworth approximation.

Another important quantity to characterize the conditional distribution
(\ref{l21}) is the third moment (\ref{l29}). For the Gaussian distribution
this moment is always equal to zero, while for the weakly nonlinear case 
a good measure of it (and the asymmetry of the distribution) is the 
quantity
\begin{equation}  \label{l31a}
    \frac{\langle(\mu - \langle\mu\rangle)^{3}\rangle_{E}}{\langle(\mu - 
    \langle\mu\rangle)^{2}\rangle_{E}^{3/2}} = \frac{\sigma \ (S_{3} - 3 
    \varrho S_{12} + 3 \varrho^{2} S_{12} - \varrho^{3} S_{3})}{(1 - 
    \varrho^{2})^{3/2}}
\end{equation}
where we have used equations (\ref{l28})-(\ref{l29}), the quantity was
expanded in powers of $\sigma$ and only the term linear in $\sigma$ was
kept.

\begin{figure}[p]  
\begin{center}
    \leavevmode
    \epsfxsize=3in
    \epsfbox[155 77 455 765]{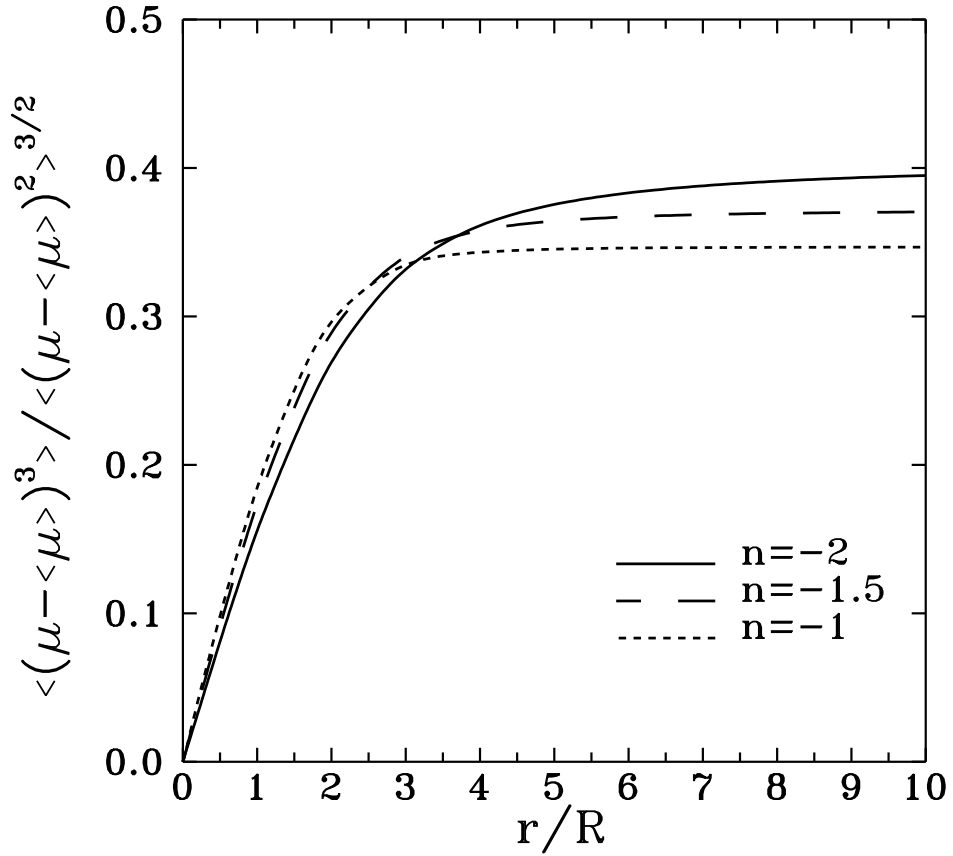}
\end{center}
    \caption{The properly normalized third moment of the conditional 
    distribution as a function of the distance from the peak, $c=r/R$, 
    for $\sigma=0.1$. Different lines show results for different power 
    spectra. At large distances the curves flatten to reach the limiting 
    values of $S_{3} \sigma$.} 
\label{sko} 
\end{figure}

Figure~\ref{sko} shows the values of the normalized third moment 
(\ref{l31a}) as a function of $c$ for $\sigma=0.1$ and different power 
spectra. It is obvious that the shape of the curves would be the same for 
any other $\sigma$. It is worth noting that the value (\ref{l31a}) which 
provides the lowest order correction to the Gaussian case does not depend 
on the height of the peak, $a$. The third moment normalized in such way 
grows with $c$ for all considered spectral indices. For large $c$ the 
values (at a given $\sigma$ and $c$) are higher for lower spectral indices 
and they approach those of the one-point Edgeworth series, $S_{3} \sigma$, 
with $S_{3}$ given by equation (\ref{e26}): 4.022, 3.714 and 3.468 
respectively for $n=-2,-1.5$ and $-1$. It is interesting that up to $c
\approx 3$ the dependence on the spectral index is weak but opposite:
the third moment grows with $n$.

\section{Application to spherical collapse}

The dynamical evolution of matter at the distance $c_{i}$ from the
peak is determined by the mean cumulative density perturbation within
$c_{i}$ which is given by
\begin{equation}  \label{l32}
    \Delta_{i}(c_{i}) = \frac{3}{c_{i}^{3}}
    \int_{0}^{c_{i}} \delta(c) c^{2} {\rm d} c
\end{equation}
where we put $\delta(c) = \sigma \langle\mu\rangle$ with 
$\langle\mu\rangle$ given by equations (\ref{l24}) and (\ref{l27}) in 
Gaussian and Edgeworth case respectively. In the linear case the function 
$\Delta_{i}(c_{i})$ is well approximated by
\begin{equation}           \label{l33}
  \Delta_{i, G}(c_{i}) = \left\{
\begin{array}{ll}
       a \sigma &\ \ {\rm for} \ \ c_{i} \ll 1 \\
       a \sigma h(n)c_{i}^{-(n+3)}   &\ \ {\rm for} \ \ c_{i} \gg 1
\end{array} \right.
\end{equation}
where $h(n)$ is a numerical factor. Its values for different spectral
indices are given in Table~\ref{h}. The approximation at large $c_{i}$ is 
accurate to within 10\% 
for $c_{i} \ge 5$. 

\begin{table}
\centering
\begin{tabular}{cll}  \hline \hline
 $n$ & \ \ \ $h$ & $S_{3} - s_{12}$ \\
  \hline
$-3.0       $& \ 1.     & 1.619   \\
$-2.5       $& \ 1.659  & 1.329   \\
$-2.0       $& \ 2.659  & 1.117   \\
$-1.5       $& \ 4.091  & 0.9757  \\
$-1.0       $& \ 6.     & 0.8966  \\
$-0.5       $& \ 8.296  & 0.8747  \\
$\ \; 0.0   $&  10.63   & 0.9064  \\
$\ \; 0.5   $&  12.27   & 0.9897  \\
$\ \; 1.0   $&  12.     & 1.124   \\
\hline
\hline
\end{tabular}
\caption{The values of the parameters $h$ and $S_{3} - s_{12}$
as functions of the spectral index $n$.}
\label{h}
\end{table}

In the weakly nonlinear case equations (\ref{l27}) and (\ref{l15a}) give
\begin{equation}           \label{l34}
  \Delta_{i, E}(c_{i}) = \left\{
\begin{array}{ll}
       a \sigma &\ \ {\rm for} \ \ c_{i} \ll 1 \\
       a \sigma h(n) \left[ 1-\sigma (S_{3}-s_{12})
       (a^{2}-1)/(2 a) \right] c_{i}^{-(n+3)}
       &\ \ {\rm for} \ \ c_{i} \gg 1
\end{array} \right.
\end{equation}
where $S_{3}$ and $s_{12}$ are given by equations (\ref{e26}) and
(\ref{l15b}) respectively. The factor $S_{3}-s_{12}$ is of the order of
unity; the exact numerical values for different $n$ are given in 
Table~\ref{h}. Equation (\ref{l34}) shows that in the limit of large 
$c_{i}$ weakly nonlinear interactions do not change the slope of the 
average initial density profile. Instead the effective height of the peak 
is changed: the profile (\ref{l34}) at large $c_{i}$ can be viewed as a 
linear profile (\ref{l33}) with the effective height of the peak
\begin{equation}  \label{l35}
    a_{eff} = a \left[ 1 - \frac{\sigma (S_{3} - s_{12})
    (a^{2}-1)}{2 a} \right].
\end{equation}

If we now assume that the matter of average density $\Delta_{i}$ within
distance $c_{i}$ from the peak collapses undisturbed onto the peak, the
spherical model can be applied. In what follows we compare the predictions
of the spherical model with Gaussian initial conditions to the ones
obtained with the initial conditions settled by the Edgeworth
approximation.

According to the spherical model the radius of maximum expansion is
related to the initial radius of the shell $r_{i}$ in the following way
\begin{equation}  \label{l36}
    r_{m} = r_{i} \frac{\Delta_{i} + 1}
    {\Delta_{i} - \delta_{c}}
\end{equation}
where $\delta_{c} = \Omega_{i}^{-1} - 1$ and $\Omega_{i}$ is the density 
parameter at some initial epoch. The maximal radius is related to the 
radius of the shell after virialization, $r$, in the following way
\begin{equation}  \label{l37}
    r = f r_{m},
\end{equation}
where $f$ is of the order of 1/2. 

From now on we will focus on the analytically tractable case of
large $c_{i}$ where the corresponding $\Delta_{i}$ are much less than
unity. Combining (\ref{l36}) and (\ref{l37}), expressing distances in
units of the smoothing scale $R$, $r = c R$, and expanding in powers of
$\Delta_{i}$ we have
\begin{equation}  \label{l38}
    \frac{c_{i}}{c} = \frac{1}{f} \frac{\Delta_{i} - \delta_{c}}
    {\Delta_{i} + 1} = \frac{1}{f} [ -\delta_{c} + (1+\delta_{c})
    \Delta_{i} + {\cal O} (\Delta_{i}^{2})].
\end{equation}
The expansion (\ref{l38}) is in fact an expansion in powers of 
$c_{i}^{-(n+3)}$ as a small parameter but it is equivalent to the 
expansion in $\sigma$ if the terms of the order of $c_{i}^{-2(n+3)}$ and
higher are neglected.

The value of $\delta_{c}$ defines the radius $c_{0}$ of the shell, the
total energy of which vanishes. All shells of radii $c_{i} < c_{0}$ are
gravitationally bound to the peak and will eventually collapse onto it.
The condition of vanishing energy leads to $\Delta_{i}(c_{0}) = \delta_{c}$ 
and in the linear case we have
\begin{equation}  \label{l39prim}
    c_{0, G} = \left[ \frac{a \sigma h(n)}{\delta_{c}} \right]^{1/(n+3)}
\end{equation}
while in the weakly nonlinear approximation
\begin{equation}  \label{l39}
    c_{0, E} = \left[ \frac{a_{eff} \sigma h(n)}{\delta_{c}}
    \right]^{1/(n+3)} = c_{0, G} \left[1 - \frac{\sigma (S_{3} -
    s_{12})(a^2 - 1)}{2 a (n+3)} \right].
\end{equation}
The weakly nonlinear $c_{0}$ is therefore decreased with respect to the 
linear one.

Since all shells which are gravitationally bound to the peak have
$\Delta_{i} (c_{i}) > \delta_{c}$ the term $\delta_{c} \Delta_{i}$ in
equation (\ref{l38}) can be neglected as being of the order of 
$\Delta_{i}^{2}$ and we have
\begin{equation}  \label{l40}
    \frac{c_{i}}{c} = \frac{\delta_{c}}{f} \left[
    \left( \frac{c_{0}}{c_{i}} \right)^{n+3} - 1 \right].
\end{equation}

The density run of the collapsed object can be estimated by assuming that
the material originally at the shell of radius $c_{i}$ ends up at $c$ so
that
\begin{equation}  \label{l41}
    \rho(c) c^{2} {\rm d} c = \rho_{i}(c_{i}) c_{i}^{2} {\rm d} c_{i}.
\end{equation}
The initial density of the shell of radius $c_{i}$ 
\begin{equation}  \label{l42}
    \rho_{i}(c_{i}) = \rho_{b,i} [1 + \delta_{i} (c_{i})],
\end{equation}
can be approximated by $\rho_{b,i}$, the background density at the initial 
epoch, since at large $c_{i}$ considered here the expected value of 
$\delta_{i}$ is very small in linear case and in fact even smaller in the
weakly nonlinear approximation as was demonstrated in equation (\ref{l27}).
The density profile is therefore in general given by
\begin{equation}  \label{l43}
    \rho(c) = \rho_{b,i} \left(\frac{\delta_{c}}{f} \right)^{3}
    \frac{[(c_{0}/c_{i})^{n+3} - 1]^{4}}{(n+4)(c_{0}/c_{i})^{n+3} - 1}.
\end{equation}
In the two limiting cases of interest that is of $c_{i}$ much smaller than
and comparable to $c_{0}$ we have
\begin{equation}  \label{l44}
    \rho(c) = \frac{\rho_{b,i}}{n+4}
    \left(\frac{\delta_{c}}{f} \right)^{3/(n+4)}
    \left( \frac{c_{0}}{c} \right)^{3(n+3)/(n+4)}
    \ \ {\rm for} \ \ c_{i} \ll c_{0}
\end{equation}
\begin{equation}  \label{l45}
    \rho(c) = \frac{\rho_{b,i}}{n+3}
    \frac{f}{\delta_{c}}
    \left( \frac{c_{0}}{c} \right)^{4}
    \ \ {\rm for} \ \ c_{i} \le c_{0}.
\end{equation}

The dependence of $\rho(c)$ on the ratio $c_{0}/c$ and the form of the 
two limiting cases are then the same in the weakly nonlinear as in the 
linear case discussed by HS. When $c_{0}$ is very 
large or very small the weakly nonlinear corrections will not affect much 
the linear results and the slopes (\ref{l44}) and (\ref{l45}) will be 
preserved. The case of $c_{0} \rightarrow \infty$ corresponds to 
$\Omega_{i} =1$. If at present we have $\Omega_{0} = 1$ then surely in the 
past also $\Omega_{i}=1$ and all profiles should have the asymptotic form 
(\ref{l44}). If the present universe is open the profile depends on the 
collapse time of the structure: the structures that collapsed later 
(with lower $\Omega_{i}$) should have smaller $c_{0}$ and steeper 
profiles because, as shown by HS by fitting numerically some $c^{- 
\alpha}$ profiles to the general formula (\ref{l43}) for given values of 
$c_{0}$, the smaller $c_{0}$, the steeper is the density profile.

The impact of weak nonlinearity can therefore be seen only at intermediate 
values of $c_{0}$ when the correction to $c_{0}$ given by equation 
(\ref{l39}) is significant. To estimate the correction to the slope 
introduced by the change in $c_{0}$ we perform fits similar to those of HS 
here first with $c_{0}$ given by the linear value (\ref{l39prim}) and then 
the weakly nonlinear one, (\ref{l39}). The fits should be treated only as 
indicative because they are not rigorous perturbatively. We adopt the 
maximum value of $a \sigma = 1$ allowed by the Edgeworth approximation and 
find that the weakly nonlinear $c_{0, E}$ (for reasonable $n \ge -2$) can 
be as low as $c_{0, G}/2$. The results of the numerical fits of the form 
$c^{-\alpha}$ to the formula (\ref{l43}) in the range $1 < c < 10$ for the 
values of $c_{0}$ corresponding to different values of $\Omega_{i} \le 1$ 
in the case of $n=-1$ and $n=-2$ are given in Table~\ref{fit}. 

The results clearly display the dependences on $\Omega_{i}$ and $n$ 
discussed by HS: profiles are steeper for lower $\Omega_{i}$ and higher 
$n$. The weakly nonlinear approximation however predicts steeper slopes 
for $\Omega_{i} < 1$ than the linear approximation does. Only the case of 
$\Omega_{i} = 1$ remains unaffected. Therefore in general weakly nonlinear 
corrections act in the same directions as decreasing $\Omega_{i}$ or 
increasing $n$. This must be taken into account were the slopes of the 
profiles used to determine these cosmological parameters. 

\begin{table} 
\centering 
\begin{tabular}{llllll}  
\hline \hline 
\ \  $n$ & \ $\Omega_{i}$ & $c_{0, G}$ & $c_{0, E}$ & $\alpha_{G}$ & 
 $\alpha_{E}$ \\    
 \hline 
 $-2$ & 0.7  &  \ 6.2      & \ 2.7      & 2.2 & 2.5 \\ 
      & 0.8  &   10.6      & \ 4.7      & 1.9 & 2.1 \\ 
      & 0.9  &   23.9      &  10.6      & 1.7 & 1.8 \\ 
      & 1.0  &  \ $\infty$ & \ $\infty$ & 1.5 & 1.5 \\ 
\hline
 $-1$ & 0.7  &  \ 3.7      & \ 2.9      & 2.8 & 2.9 \\ 
      & 0.8  &  \ 4.9      & \ 3.8      & 2.5 & 2.6 \\ 
      & 0.9  &  \ 7.3      & \ 5.7      & 2.2 & 2.3 \\ 
      & 1.0  &  \ $\infty$ & \ $\infty$ & 2.0 & 2.0 \\ 
\hline \hline 
\end{tabular} 
\caption{The slopes of profiles $c^{-\alpha}$ fitted in the range $1 < c 
< 10$ in the linear (G) and weakly nonlinear (E) case for spectral indices 
$n=-2$ and $n=-1$ and different values of $\Omega_{i}$.} 
\label{fit} 
\end{table}

Another important quantity characterizing the protoobject is its mass. We 
assume that the mass contained initially within radius $c_{0}$  
\begin{equation}  \label{l46}                                             
    M(c_{0}) = \frac{4 \pi}{3} (c_{0} R)^{3} \rho_{b,i} 
    (1 + \Delta_{i}(c_{0})) 
\end{equation} 
does not change as the shells collapse and may be treated as the final 
mass of the protoobject. In this case the quantitative prediction 
concerning the correction to the mass gathered by the peak can be made. 
In the linear approximation the mass $M_{G}$ is calculated with $c_{0, G}$ 
and $\Delta_{i, G}(c_{0})$ given by equations (\ref{l39prim}) and 
(\ref{l33}) respectively while in the weakly nonlinear case ($M_{E}$) 
they should be replaced by the values given in equations (\ref{l39}) and 
(\ref{l34}). For the ratio of the masses we obtain 
\begin{equation}  \label{l47} 
    \frac{M_E}{M_G} = 1 - 3 \frac{(S_{3}-s_{12})(a^2-1)}{2 (n+3) a} 
    \sigma.   
\end{equation}
Thus the mass gathered by the peak predicted by weakly nonlinear 
approximation is always smaller than in the linear case and the 
correction is larger for lower spectral indices.

\section{Concluding remarks}

The picture emerging from the presented results is the following. In
the weakly nonlinear field the average overdensity around the peak, which
drives the evolution of matter around it, behaves as if produced by the
peak of reduced height embedded in the linear field. Weakly nonlinear
interactions decrease the size of region from which the peak gathers mass
and the slope of the final density profile is steepened. The type of 
dependence of the profiles on the cosmological parameters $\Omega$ and $n$ 
derived by HS is preserved.

It should be remembered that the results presented here were obtained for
the mean density perturbation, $\langle\Delta_{i, E}\rangle$, using the 
mean normalized density contrast $\langle\mu\rangle$ calculated from the 
conditional distribution (\ref{l21}). In order to specify the uncertainty 
of the results it would be desirable to know also the standard deviation 
of $\Delta_{i, E}$, $\sigma_{\Delta}$, since only below the scale 
determined by the condition $\Delta_{i, E}=\sigma_{\Delta}$ the 
perturbation $\Delta_{i, E}$ is significant. Such scale would always be 
finite and dependent on the power spectrum. In the case of $\Omega_{0}=1$ 
universe this would mean that we cannot indiscriminately apply the limit 
$c_{0} \rightarrow \infty$, which is rather unrealistic: gravitational 
influence of a peak cannot reach infinite distances because there are 
always neighbouring peaks that gather mass. Calculating $\sigma_{\Delta}$ 
in the weakly nonlinear regime would however require constructing the 
three-point Edgeworth series, which is beyond the framework of this 
paper.

There are other limitations to the weakly nonlinear approach to the 
collapse of peaks presented here. The most obvious one is the condition $a 
\sigma < 1$ that must be satisfied for the Edgeworth approximation to be 
valid. Is the evolution in this regime representative of the fully 
nonlinear clustering? Another question is how the linear and weakly 
nonlinear results should be compared. Throughout this paper we assumed 
the linear value of the variance of the density which is correct as far as 
we use the lowest order Edgeworth approximation. However, \L okas et al. 
(1996) have shown that the value of $\sigma$, the typical fluctuation 
of density, itself changes during weakly nonlinear evolution. It remains 
to be understood how the evolution of a typical fluctuation and evolution 
of matter around a peak are related.

Observational data concerning the halo density profiles are presently 
very sparse and derived most often from the fact that the observed 
rotation curves of galaxies are flat, which in the Einstein-de~Sitter 
universe leads to the profile $r^{-2}$ that according to formula 
(\ref{l44}) corresponds to $n=-1$. Such spectral index is roughly 
consistent with what is observed at weakly nonlinear scales but we can 
hardly go beyond such rough consistency checks as far as observations are 
concerned. Therefore although such data are of potential cosmological 
interest the only way to verify the theoretical predictions is to resort 
to N-body simulations.

The profiles of dark halos have been measured in N-body simulations by 
many authors including Dekel, Kowitt \& Shaham (1981) Quinn, Salmon \& 
Zurek (1986), Efstathiou et al. (1988) and Crone, Evrard \& Richstone 
(1994). The results of such simulations seem to converge to a statement 
that the present structure of halos retains information on the initial 
conditions and displays the dependence on cosmological parameters. Crone 
et al. (1994) performed simulations for power-law spectra and 
different density parameters $\Omega_{0}$ and confirmed the overall trend 
of steeper density profiles with increasing $n$ or decreasing $\Omega$ as 
predicted by HS. However, the fitted slopes were 
systematically steeper than those given by HS. As discussed in the 
previous section only in the case of the open universe such discrepancy can
be assigned to the weakly nonlinear effects. The detailed comparison of the 
predictions of perturbation theory with the results of N-body simulations 
requires further assumptions (concerning e.g. the epoch of the formation 
of objects, the most realistic value of $a \sigma$ to be adopted) and is 
currently under way. 

Recently Sheth \& Jain (1997) proposed a new derivation of the slopes 
of halo profiles from the shape of nonlinear correlation function based on 
the assumption that the halo-halo correlations can be neglected i.e. that 
for sufficiently small scales the input to the correlation function comes 
mostly from one halo. The slope they obtain, $\alpha = 3 (n+4)/(n+5)$, can 
be derived within the formalism of HS, and would be obtained instead of 
the result (\ref{l44}), if the initial profile of the cumulative density 
is chosen to be $c_{i}^{-(n+4)}$ instead of $c_{i}^{-(n+3)}$ as in 
equations (\ref{l33})-(\ref{l34}). Sheth \& Jain (1997) claim that such an 
initial profile is well motivated by the results of Bardeen et al. (1986) 
who found that as the peak height decreases the initial profile becomes 
steeper than $c_{i}^{-(n+3)}$. In this case the approximation used here 
(i.e. regions of high enough density, and not necessarily peaks, are the 
progenitors of structure) would be less accurate. 

This objection can only be valid under condition that smaller peaks can 
collapse by themselves to form objects. N-body simulations performed 
by Katz, Quinn \& Gelb (1993) and van de Weygaert \& Babul (1994) suggest 
the opposite: there is no clear correspondence between peaks in the 
initial density field and the actual halos for smaller peaks. 
For example, Katz et al. (1993) find that most of the peaks expected to 
form cluster-size objects ($a \ge 4$) indeed end up in such objects while  
the peaks from which the galaxy-size objects form ($a \approx 2$) may end 
up in a galaxy-size group as well as merge with a larger object. Even more 
possibilities for the history of a peak, including its breaking into few 
distinct halos, were found by van de Weygaert \& Babul (1994).

\section*{Acknowledgements}

I wish to thank Roman Juszkiewicz for suggesting to me the investigation 
of the two-point Edgeworth series. I am also grateful to Micha\l \ 
Chodorowski and Fran{\c c}ois Bouchet for their comments at different 
stages of this work. The hospitality of the Institut d'Astrophysique de 
Paris, where part of this work was done, is kindly acknowledged. This work 
was supported in part by the Polish State Committee for Scientific 
Research grants No. 2P30401607, 2P03D00310 and 2P03D01313, and the French 
Ministry of Research and Technology within the programme Jumelage 
(Astronomie Pologne).

\newpage 

\section*{References}

\begin{description}

\item Adler, R. J. 1981,
    ``The Geometry of Random Fields" (Chichester: Wiley)
\item Bardeen, J. M., Bond, J. R., Kaiser, N. \& Szalay, A. S. 1986,
    ApJ, 304, 15
\item Bernardeau, F. 1994,
   ApJ, 433, 1
\item Bernardeau, F. 1996, A\&A, 312, 11
\item Bernardeau, F. \& Kofman, L. 1995,
    ApJ, 443, 479
\item Crone, M. M., Evrard, A. E. \& Richstone, D. O. 1994, 
    ApJ, 434, 402
\item Dekel, A. 1981, 
    A\&A, 101, 79
\item Dekel, A., Kowitt, M. \& Shaham, J. 1981, 
    ApJ, 250, 561
\item Efstathiou, G., Frenk, C. S., White, S. D. M. \& Davis, M. 1988,
    MNRAS, 235, 715
\item Gott, J. R. 1975,
    ApJ, 201, 296
\item Gunn, J. E. 1977,
    ApJ, 218, 592
\item Gunn, J. E. \& Gott, J. R. 1972,
    ApJ, 176, 1
\item Hoffman, Y. \& Shaham, J. 1985,
    ApJ, 297, 16 (HS)
\item Juszkiewicz, R., Bouchet, F. R. \& Colombi, S. 1993,
    ApJL, 412, L9
\item Juszkiewicz, R., Weinberg, D. H., Amsterdamski, P., Chodorowski, M.
   \& Bouchet, F. R. 1995,
   ApJ, 442, 39
\item Katz, N., Quinn, T. \& Gelb, J. M. 1993,
    MNRAS, 265, 689
\item \L okas, E. L., Juszkiewicz, R., Weinberg, D. H. \& Bouchet, F. R.
   1995,
   MNRAS, 274, 730
\item \L okas, E. L., Juszkiewicz, R., Bouchet, F. R. \& Hivon, E. 1996,
   ApJ, 467, 1
\item Longuet-Higgins, M. S. 1963,
   Journal of Fluid Mechanics, 17, 459
\item Longuet-Higgins, M. S. 1964,
   Radio Science Journal of Research, 68D, 1049
\item Peebles, P. J. E. 1980,
    ``The Large-scale Structure of the Universe"
    (Princeton: Princeton University Press)
\item Peebles, P. J. E. 1984,
    ApJ, 277, 470
\item Quinn, P. J., Salmon, J. K. \& Zurek, W. H. 1986, 
    Nature, 322, 329
\item Sheth, R. K. \& Jain, B. 1997, 
    MNRAS, 285, 231
\item van de Weygaert, R. \& Babul, A. 1994,
    ApJ, 425, L59

\end{description}

\end{document}